\documentclass[11pt]{article}
\usepackage{amsmath,amssymb,color,graphics,epsfig,tikz}

\usepackage{amsfonts}

\newcommand{\hoch}[1]{$\, ^{#1}$}

\newcommand{\be}{\begin{equation}}
\newcommand{\ee}{\end{equation}}
\newcommand{\bea}{\setlength\arraycolsep{2pt} \begin{eqnarray}}
\newcommand{\eea}{\end{eqnarray}}
\newcommand{\nn}{\nonumber}

\def\ft#1#2{{\textstyle{\frac{\scriptstyle #1}{\scriptstyle #2} } }}
\def\fft#1#2{{\frac{#1}{#2}}}

\def\0{{\sst{(0)}}}
\def\1{{\sst{(1)}}}
\def\2{{\sst{(2)}}}
\def\3{{\sst{(3)}}}
\def\4{{\sst{(4)}}}
\def\5{{\sst{(5)}}}
\def\6{{\sst{(6)}}}
\def\7{{\sst{(7)}}}
\def\8{{\sst{(8)}}}
\def\sst#1{{\scriptscriptstyle #1}}

\def\ep{{\epsilon}}
\def\del{{\partial}}

\def\im{{\rm i}}

\begin{document}

\begin{flushright}
\hfill{MIFPA-13-11}
\end{flushright}

\vspace{25pt}
\begin{center}
{\large {\bf Not Conformally-Einstein Metrics in Conformal Gravity}}

\vspace{10pt}
Hai-Shan Liu\hoch{1},  H. L\"u\hoch{2}, C.N. Pope\hoch{3,4} and J.F. V\'azquez-Poritz\hoch{5}

\vspace{10pt}

\hoch{1} {\it Institute for Advanced Physics \& Mathematics,\\
Zhejiang University of Technology, Hangzhou 310023, China}

\vspace{10pt}

\hoch{2}{\it Department of Physics, Beijing Normal University,
Beijing 100875, China}

\vspace{10pt}

\hoch{3} {\it George P. \& Cynthia Woods Mitchell  Institute
for Fundamental Physics and Astronomy,\\
Texas A\&M University, College Station, TX 77843, USA}

\vspace{10pt}

\hoch{4}{\it DAMTP, Centre for Mathematical Sciences,
 Cambridge University,\\  Wilberforce Road, Cambridge CB3 OWA, UK}

\vspace{10pt}

\hoch{5}{\it New York City College of Technology, The City
University of New York\\ 300 Jay Street, Brooklyn NY 11201, USA}

\vspace{40pt}

\underline{ABSTRACT}
\end{center}

The equations of motion of four-dimensional conformal gravity, whose Lagrangian is the square of the Weyl tensor, require that the Bach tensor $E_{\mu\nu}=
(\nabla^\rho\nabla^\sigma + \ft12 R^{\rho\sigma})C_{\mu\rho\nu\sigma}$
vanishes.  Since $E_{\mu\nu}$ is zero for any Einstein metric,
and any conformal scaling of such a metric, it follows that large classes of
solutions in four-dimensional conformal gravity are simply given by metrics
that are conformal to Einstein metrics (including Ricci-flat).
In fact it becomes more intriguing to find solutions that are {\it not}
conformally Einstein.  We obtain five new such vacua, which are homogeneous
and have asymptotic generalized Lifshitz anisotropic scaling symmetry.
Four of these solutions can be further generalized to metrics that are
conformal to classes of pp-waves, with a covariantly-constant null vector.
We also obtain large classes of generalized Lifshitz vacua in
Einstein-Weyl gravity.

\thispagestyle{empty}

\pagebreak



\newpage

\section{Introduction}

Einstein's theory of gravity can be further extended by adding higher-order
curvature terms, without violating the underlying principles of General
Relativity.  Such terms also arise naturally as
perturbations to the low-energy effective actions of string theories.
Because the equations of motion now involve higher derivatives and
a higher degree of non-linearity, much less is known about the solutions
in higher-derivative gravities.  In four dimensions, since the
Gauss-Bonnet quadratic curvature invariant is a total derivative,
the possible independent quadratic curvature modifications are equivalent just
to $R^2$ and $R^{\mu\nu}R_{\mu\nu}$. This implies that
Einstein metrics, including Schwarzschild-AdS or Kerr-AdS metrics,
are automatically solutions in cosmological gravities extended with
quadratic curvature terms.  It is an interesting, although generally
rather hard, problem to find
new solutions that are not merely Einstein metrics.

       In this paper, we shall first consider solutions of purely
conformal gravity in four dimensions, for which the Lagrangian can be written
as
\begin{equation}
{\cal L}= \ft12 \sqrt{-g} \alpha C^{\mu\nu\rho\sigma} C_{\mu\nu\rho\sigma}\,,
\end{equation}
where $C_{\mu\nu\rho\sigma}$ is the Weyl tensor.  The equations of motion
following from this Lagrangian imply that the Bach tensor
\be
E_{\mu\nu}= \Big(\nabla^\rho\nabla^\sigma +\ft12 R^{\rho\sigma}\Big)
C_{\mu\rho\nu\sigma}
\ee
must vanish.

    The most general
spherically-symmetric black hole solution in conformal gravity, up
to an overall conformal factor, was obtained in \cite{Riegert:1984zz}. It
turns out that it is locally conformal to the Schwarzschild-AdS black hole
\cite{Lu:2012xu}; however, since the conformal factor can be singular at
infinity these black holes are globally distinct from Schwarzschild-AdS.
(Charged static solutions with toroidal and hyperbolic topologies were given
in \cite{Lu:2012ag,Li:2012gh}, where the AdS/CFT applications to Fermi and
non-Fermi liquids were studied.)

    The most general metrics in conformal gravity within the Plebanski
class have also been constructed, in \cite{Mannheim:1990ya}. It was shown in
\cite{Liu:2012xn} that these again include new (rotating) black holes that
are globally distinct from the usual Kerr-AdS solutions.  The general metric
obtained in \cite{Mannheim:1990ya} is conformal to the
Plebanski-Demianski metric, which is the most general type-D Einstein metric.

    This leads one to wonder about the existence of solutions in conformal
gravity that are not conformal to Einstein metrics.  One such solution was
indeed constructed in \cite{Nurowski:2000cq}.\footnote{It is 
interesting to note that, as was shown recently in \cite{djw},
a partially massless spin-2 field does not propagate in a solution 
to conformal gravity unless it is conformally Einstein.}  
One goal of the present paper
is to find more metrics in conformal gravity that are not conformal to
Ricci-flat or Einstein metrics.  (In this paper, when we refer to
Einstein metrics, we shall include Ricci-flat metrics as well.)  It
is not necessarily simple to test whether two ostensibly different
solutions are actually conformal to each other.  For this reason, we focus
first on classes of metric ans\"atze that are homogeneous with a
generalized Lifshitz scaling symmetry.  Because of the homogeneity and
the symmetries, it is evident in these cases that such solutions are not
conformally related to each other.

    In section 2, we review a condition derived in \cite{Gover:2004ar} for
testing
whether a solution is conformally Einstein.  It provides a necessary,
but not sufficient, criterion for a metric to be conformal to an
Einstein metric.  Thus, if this condition is not obeyed for a given metric, then the metric
cannot be conformally transformed to be Einstein. We use the gyrating
vacua \cite{Lu:2012cz} in conformal gravity as examples to demonstrate that
for appropriate choices of the parameters, the metrics can be either (I)
Einstein, or (II) conformally Ricci-flat, or (III) not conformal to any
Einstein metric.

     In section 3, we consider a large class of homogeneous metrics with
generalized Lifshitz scaling symmetry.  By demanding that these metrics
satisfy the equations of motion of conformal gravity, we obtain six
solutions that are not conformally-Einstein, including one that is conformal
to the previously-known metric \cite{Nurowski:2000cq}.  We argue that
these solutions are distinct, in the sense that they are also not conformal
to each other.  We find that four solutions can be viewed as conformal to
certain pp-wave metrics (whose defining property is that they admit
a covariantly-constant null vector).

     In section 4, we show that the four solutions mentioned above can be
generalized to much broader classes of inhomogeneous pp-waves in
conformal gravity.
In section 5, we consider cohomogeneity-one Bianchi IX metrics in
conformal gravity.  Whilst there exist triaxial solutions that are not
conformally Einstein, all the biaxial solutions are conformally Einstein. We
argue that this suggests that it is unlikely that there will exist
any rotating black holes in conformal gravity beyond those that are
already known, which are conformally Einstein.

     In section 6, we consider the more general Einstein-Weyl theory of
gravity, for which the Lagrangian is Einstein-Hilbert with a
cosmological constant in addition to the Weyl-squared term.
We obtain some solutions that are homogeneous metrics with generalized
Lifshitz scaling symmetries.  This paper ends with conclusions in section 7.

\section{Testing of conformally-Einstein metrics}

It was shown in \cite{Gover:2004ar} that if a metric $ds^2$ is conformal to an Einstein metric $d\tilde s^2$ with $ds^2=\Omega^{-2} d\tilde s^2$, then one must have
\begin{equation}
\nabla_\mu C^{\mu\nu\rho\sigma} - (D-3) V_\mu C^{\mu\nu\rho\sigma}=0\,,\label{conformalcon}
\end{equation}
for some appropriate $V_\mu$.  The conformal factor $\Omega$
can then be obtained by
$V_\mu =\partial_\mu \log\Omega$.  This was used recently to distinguish
the spherically-symmetric solutions in some six-dimensional conformal
gravity theories that are conformally Einstein from those that are not \cite{Lu:2013hx}.

The existence of a solution for the vector $V_\mu$ is a necessary condition
for the metric to be Einstein.  Thus, if we have a metric for which there is
no such solution, then the metric is not conformally Einstein.  On the other hand, if there
exists a solution for $V_\mu$ then it is not guaranteed that the
metric is conformally Einstein, and it is still necessary to find the
conformal factor to check whether the metric is indeed
conformally Einstein or not.

Let us consider the gyrating homogeneous vacua ansatz proposed
in \cite{Lu:2012cz}:
\begin{equation}
ds^2 = \ell^2 \Big( \fft{dr^2 - 2 du dv + dx^2}{r^2} +
\fft{2c_1 du dx}{r^{z+1}} -\fft{c_2\,du^2}{r^{2z}}\Big)\,,\label{gyraton}
\end{equation}
where $c_1$ and $c_2$ are constants.  The metric is AdS when $z=1$ and/or $c_1=0=c_2$.  For non-vanishing $c_1$, aside from the AdS metric, there are three solutions in conformal gravity:
\begin{eqnarray}
{\rm I}:&& z=-2\,,\qquad c_2=-\ft12c_1^2\,;\cr
{\rm II}:&& z=0\,,\qquad\ \ \hbox{$c_i$ arbitrary}\,;\cr
{\rm III}:&& z=-1\,,\qquad c_2=-\ft12c_1^2\,.
\end{eqnarray}
If we substitute the metric I into (\ref{conformalcon}),
we find that $V_\mu=0$. Indeed, it is easy to verify that the metric I
is already Einstein, with $R_{\mu\nu} = -(3/\ell^2) g_{\mu\nu}$.

If we substitute metric II into (\ref{conformalcon}), we find that
\begin{equation}
V_r=\fft{1}{r}\,,\qquad V_v=0=V_x\,,
\end{equation}
and that there is no constraint on $V_u$. This implies that the conformal
factor is $\Omega^2 = r^2 f(u)^2$.  Since condition
(\ref{conformalcon}) is necessary but not sufficient, it is still
necessary to find a solution for
the function $f(u)$ that makes the conformally-transformed metric
Einstein.  In fact, the metric becomes Ricci flat, if $f$ is taken to be
\begin{equation}
f(u)=\fft{1}{\cos(\ft12\sqrt{c_1^2 + 2 c_2}\, u)}\,.
\end{equation}
The Ricci-flat metric can be cast in the form
\begin{equation}
ds^2 = \cos^2\theta \Big(dr^2 + (2c_2-c_1^2) dx^2\Big) + \fft{4 c_2 r^2}{2c_2-c_1^2}\,d\theta^2 + 4 \cos\theta\, d\theta (dv + c_1 r\, dx)\,.
\end{equation}
The metric is of the pp-wave type, which we define to be the one with a
covariantly-constant null vector $\ell=\partial_v$, which also happens to be a
Killing vector in this example.  Note that all the curvature polynomials
vanish identically.

For the metric III, there is no solution of $V_\mu$, and hence the metric cannot be conformally transformed to any Einstein metric.  The construction and the classification of these solutions go beyond Einstein metrics. One focus of this paper is to construct further examples of such solutions.

So far we have considered the gyratons with $c_1$ non-vanishing.  The solution (\ref{gyraton}) with $c_1=0$ belongs to the pp-wave type of Schr\"odinger metrics \cite{Son:2008ye,Balasubramanian:2008dm}.  In conformal gravity, we have $z=\pm \fft12$.  The $z=-\fft12$ solution is the Kaigorodov metric \cite{kaig} which is Einstein, whilst the $z=+\fft12$ solution is conformal to a Ricci-flat pp-wave.

\section{Homogeneous and not conformally-Einstein vacua}

\subsection{The metrics and their properties}

In this section, we consider the metric ansatz
\be\label{metric}
ds^2=\fft{dr^2}{r^2}+\sum_{i=k}^3 d_k r^{2z_k} dx_k^2+
  \ft12 \sum_{i\ne j\ne k} \sqrt{c_k} r^{z_i+z_j} dx_i dx_j\,.
\ee
Here $z_k$, $d_k$ and $c_k$ are constant parameters. Note that if all
$z_k$'s are equal, then the metric becomes the usual local patch of AdS, which we shall not
consider further.
The metrics are all homogeneous, with the generalized Lifshitz scaling
symmetry
\begin{equation}
r\rightarrow \lambda r\,,\qquad x_i\rightarrow
   \lambda ^{-z_i} x_i\,,\qquad i=1,2,3\,.
\end{equation}
The components of the curvature tensor in the vielbein base
\begin{equation}
e^{\bar r} = \fft{dr}{r}\,,\qquad e^i=r^{z_i} dx_i
\end{equation}
are all constants, independent of $r$.  It is clear that the metric ansatz
contains the Lifshitz \cite{korlib,kacliumul},
Schr\"odinger \cite{Son:2008ye,Balasubramanian:2008dm} and gyraton
\cite{Lu:2012cz} solutions as special cases. In Einstein gravity, the
Lifshitz and Schr\"odinger solutions require some appropriate matter
energy-momentum tensor, whilst they can arise naturally in
higher-derivative gravity theories without additional matter \cite{Lu:2012xu}.

Three classes of solutions can arise:  (I) The metric (\ref{metric}) is
Einstein, with non-zero cosmological constant; (II) It is conformal to an
Einstein metric (which is in fact Ricci flat in this case); or (III) It is not conformal to any Einstein metric.
In this paper, we shall be concerned with only the third class of solutions,
since the study of solutions in the first and second classes reduces
to a study of solutions in Einstein gravity.

For the diagonal case, all $c_k$ vanish and the $d_k$ can be set to unity by
rescaling the $x_k$. After a Wick rotation, the metric becomes
\be\label{diag-metric}
ds^2=-r^{2z_3}dt^2+\fft{dr^2}{r^2}+r^{2z_1} dx_1^2+r^{2z_2} dx_2^2\,.
\ee
Metrics of this kind, with  multiple anisotropic scaling symmetries,
were constructed in Einstein gravity coupled to multiple massive vectors
\cite{Taylor:2008tg}.
The metric (\ref{diag-metric}) is a solution in conformal gravity if
\begin{equation}
z_1^2 + z_2^2 + z_3^2 - 2 z_1 z_2 - 2 z_1 z_3 - 2 z_2 z_3=0\,.
\end{equation}
It turns out that these metrics can be rendered Ricci-flat by means of a scaling with the conformal factor
\be
\Omega^2=r^{-z_1-z_2-z_3}.
\ee
Thus, any solution of the form (\ref{metric}) that is not conformally Einstein
necessarily involves at least one off-diagonal term.  We find six such
solutions.\footnote{Non-conformally Einstein metrics with vanishing Bach tensor
were discussed in the context of diagonal Bianchi I models in
\cite{schmidt}.}  
In presenting these solutions, we have performed appropriate
analytical continuation so that the metrics are all real and of
Lorentzian signature.

\bigskip
\noindent{\bf Solution 1:}
\begin{equation}
ds^2=\fft{dr^2}{r^2} + r^2 ( 2 du dv+dx^2) + r^5 dv^2 -
           \fft{5 du^2}{r}\,.
\end{equation}
The metric has the Ricci tensor $R_{ij}=\lambda_{ij} g_{ij}$ (no sum) where $\lambda_{ij}$ are:
\begin{equation}
\lambda_{rr}=-\ft{27}{4}\,,\qquad \lambda_{uu}=\ft34\,,\qquad
\lambda_{vv}=-\ft{33}{4}\,,\qquad\lambda_{xx}=-3\,,\qquad \lambda_{uv}=\ft34\,.
\end{equation}
Using the criteria for determining the Petrov type that we summarise in
appendix A, we find that this metric is of Petrov Type I.

\bigskip
\noindent{\bf Solution 2:}
\begin{equation}
ds^2 = \fft{dr^2}{r^2} + \fft{2 du dv}{r}+ \fft{dx^2}{r^4}+r^2 du^2\,.
\end{equation}
with
\begin{equation}
\lambda_{rr}=-\ft92\,,\qquad \lambda_{uu}=-\ft32\,,\qquad
\lambda_{xx}=-6\,,\qquad \lambda_{uv}=-\ft32\,.
\end{equation}
It is of Petrov Type II.

\bigskip
\noindent{\bf Solution 3:}
\be
ds^2 = \fft{dr^2}{r^2}+\fft{2 du dv}{r}+dx^2+ 6 r \, du dx+
          7 r^2 du^2\,,
\ee
with
\begin{equation}
\lambda_{rr}=\lambda_{ux}=\lambda_{uv}=-\ft{1}{2}\,,\qquad \lambda_{uu}=-\ft{13}{14}\,,\qquad
\lambda_{xx}=0\,.
\end{equation}
It is of Petrov Type III.

\bigskip
\noindent{\bf Solution 4:}
\be
ds^2 = \fft{dr^2}{r^2}+\fft{2 du dv}{r}+\fft{dx^2}{r^4}+
   6 r^2 du dx+ 2 r^8 du^2\,,
\ee
with
\begin{equation}
\lambda_{rr}=-\ft{9}{2}\,, \qquad \lambda_{uu}=-\ft{33}{4}\,,\qquad \lambda_{xx}=\lambda_{ux}=-6\,,\qquad \lambda_{uv}=-\ft{3}{2}\,.
\end{equation}
It is of Petrov Type II.

\bigskip
\noindent{\bf Solution 5:} The fifth solution is a gyrating Schr\"odinger geometry discussed in section 2 (with $z=-1$, $c_1=\sqrt{2}$ and $c_2=-1$ in the gyrating Schr\"odinger ansatz):
\be
ds^2 = \fft{dr^2+ 2 du dv+dx^2}{r^2}+ 4 du dx+  2 r^2 du^2\,,
\ee
with
\begin{equation}
\lambda_{rr}=\lambda_{xx}=\lambda_{uv}=-3\,,\qquad \lambda_{uu}=-1\,,\qquad \lambda_{ux}=-2\,.
\end{equation}
It is of Petrov Type III.

\bigskip
\noindent{\bf Solution 6:}
\begin{equation}
ds^2 = \fft{dr^2}{r^2} + 10 du \Big(\fft{dv}{r^2} -\fft{2dx}{r}\Big) +
   2 r dx dv + \fft{10 du^2}{r^4} + r^2 dx^2\,.
\end{equation}
It is of Petrov Type N. This solution is conformal to the one obtained in
\cite{Nurowski:2000cq}, with the conformal factor $r^2$.

Having obtained these six solutions, and determined that they are not
conformally-Einstein, it is natural to ask whether these solutions are
conformal to each other.  Since the Petrov classification is conformally
invariant, the only possible candidates for being conformally related are
solutions 2 and 4, or else solutions 3 and 5.
It is clear that if the putative conformal factor in either of these
cases is a function of $r$ only, then it will break the scaling symmetry.  On the other hand, if
the conformal factor involves the
coordinates of $u,v,x$ as well, then constant shift symmetries along these
directions will be broken.  Thus, we expect that these solutions are not
conformal to each other.

The properties of the six solutions are summarized by the following table.
\vspace{.5cm}
\begin{center}
\begin{tabular}{|c|c|c|c|c|c|c|}
\hline solution & $R_{\mu\nu}R^{\mu\nu}/R^2$ & Riem$^2/R^2$ & Riem$^3_{(1)}/R^3$ &  Riem$^3_{(2)}/R^3$ &Petrov Class\\
\hline\hline 1 & $\fft{179}{441}$ & $\fft{713}{1323}$ & $-\fft{24,229}{83,349}$ & $\fft{5573}{333,396}$ & type I\\
\hline 2  & $\fft13$ & $\fft{97}{243}$ & $-\fft{1409}{6561}$ & $\fft{145}{26,244}$ &type II\\
\hline 3  & $\fft13$ & $\fft13$ & $-\fft19$ & $\fft{1}{36}$ &type III \\
\hline 4  & $\fft13$ & $\fft{97}{243}$ & $-\fft{1409}{6561}$ & $\fft{145}{26,244}$ &type II\\
\hline 5  & $\fft14$ & $\fft16$ & $-\fft{1}{36}$ & $\fft{1}{72}$ &type III\\
\hline 6 & $\fft{13}{27}$ & $\fft{17}{27}$ & $-\fft{13}{9}$ & $-\fft{17}{108}$ & type N\\
\hline
\end{tabular}
\end{center}
\medskip
{\footnotesize Table 1: The curvatures and Petrov types of the six homogeneous and not conformally-Einstein vacua.}

\subsection{PP-waves in conformal gravity}

The solutions 2-6, unlike solution 1, all admit a global null vector
$\ell=\partial_v$.  However, it is not covariantly constant for any of these
solutions.  In fact we find no covariantly-constant null vector for any of
our metrics, and hence none of them are of the pp-wave type.  However,
solutions 2, 3, 4 and 5 can be transformed into pp-waves by multiplying
the metric by a particular conformal factor that depends on the $r$
coordinate.  With the appropriate conformal factors, the metrics 2, 3, 4 and 5
all then admit a global null vector $V=\partial_v$ that is also covariantly
constant.  This null vector also happens to be a Killing vector.  Of course,
as we have established, these are pp-waves of conformal gravity which are neither
Ricci-flat nor Einstein.  Here we list the corresponding pp-waves and
present some of their properties.  Their Petrov classification is the same as the corresponding homogeneous vacuum solution.

\medskip
\noindent{\bf PP-wave 1}: The solution is conformal to the homogeneous solution 2, with a conformal factor $r$, namely
\begin{equation}
ds^2 = \fft{dr^2}{r} + 2 du dv+ \fft{dx^2}{r^3}+r^3 du^2\,.
\end{equation}
The Ricci-tensor becomes diagonal in the coordinate base with the following components:
\begin{equation}
R_{rr}=-\fft{3}{r} g_{rr}\,,\qquad R_{uu} = -\fft{3}{2r} g_{uu}\,,\qquad R_{xx}=-\fft{3}{r} g_{xx}\,.
\end{equation}
We find that the Ricci scalar is $R=-6/r$ and that the other 
non-vanishing curvature invariants are
\begin{equation}
R_{\mu\nu}R^{\mu\nu}=\ft12 R^2\,,\qquad
{\rm Riem}^2=R^2\,,\qquad
{\rm Riem}^3_{(1)}=-R^3\,,\qquad {\rm Riem}^3_{(2)}=0\,.
\end{equation}
It is clear that the metric has a curvature singularity at $r=0$.
\medskip

\noindent{\bf PP-wave 2}: This is conformal to the homogeneous solution 3, with a conformal factor $r$,  namely
\be
ds^2 = \fft{dr^2}{r}+2 du dv+dx^2+ 6 r^2 \, du dx+
          7 r^3 du^2\,.
\ee
All the curvature polynomials vanish, but with the following non-vanishing Ricci-tensor components
\begin{equation}
R_{uu} = -\fft{27}{14 r}\, g_{uu}\,,\qquad R_{ux} = -\fft{1}{r}\, g_{ux}\,.
\end{equation}
Thus the metric has a curvature singularity at $r=0$.
\medskip

\noindent{\bf PP-wave 3}: This is conformal to the homogeneous solution 4, with a conformal factor $r$, {\it i.e.}
\be
ds^2 = \fft{dr^2}{r}+2 du dv+\fft{dx^2}{r^3}+
   6 r^3 du dx+ 2 r^9 du^2\,.
\ee
The non-vanishing Ricci-tensor components are
\begin{equation}
R_{rr}=-\fft{3}{r} g_{rr}\,,\quad R_{uu}=-\fft{45}{4r} g_{uu}\,,\quad
R_{xx}=-\fft{3}{r} g_{xx}\,,\qquad R_{ux}=-\fft{6}{r} g_{ux}
\,.
\end{equation}
Thus we have $R=-6/r$, together with
\begin{equation}
R_{\mu\nu}R^{\mu\nu}=\ft12 R^2\,,\qquad
{\rm Riem}^2=R^2\,,\qquad
{\rm Riem}^3_{(1)}=-R^3\,,\qquad {\rm Riem}^3_{(2)}=0\,,
\end{equation}
\medskip
and there is a curvature singularity at $r=0$.

\noindent{\bf PP-wave 4}: This is conformal to the homogeneous solution 5, but now with a conformal factor $r^2$.  In other words, we have
\be
ds^2 = dr^2+ 2 du dv+dx^2+ 4 r^2 du dx+  2 r^4 du^2\,.
\ee
The non-vanishing Ricci-tensor components are given by
\begin{equation}
R_{uu}=-\fft{2}{r^2} g_{uu}\,,\qquad R_{ux}=-\fft{1}{r^2} g_{ux}\,.
\end{equation}
However, all the curvature polynomials vanish identically.  The metric is nevertheless singular at $r=0$.

\medskip
    It is worth pointing out that the Schr\"odinger solution also belongs
to the pp-wave class of metrics.  However, it is homogeneous whilst the pp-waves we
obtained above are of cohomogeneity one.

\section{More general classes of not conformally-Einstein metrics}

As in the case of the Schr\"odinger solution which is a special case of a more general pp-wave solution, we show that the homogeneous vacuum solutions 2-5 can be further generalized to become much more general non-homogeneous metrics.  These metrics can be viewed as flowing from one vacuum in $r=0$ to another in $r=\infty$.  We consider the conditions such that the solutions are not conformally-Einstein. It turns out that all of these solutions are conformal to pp-waves with a covariantly-constant null vector.

\bigskip
\noindent{\bf Generalising solution 2:}
\be
ds^2 =\fft{dr^2}{r^2} + \fft{2 du dv}{r}+ \fft{dx^2}{r^4}+r^2 h du^2\,.
\ee
where
\be
h = c_0\log r+c_1+\fft{c_2\log r}{r^3}+\fft{c_3}{r^3}\,.
\ee
Note that the $c_i$'s can be arbitrary functions of $u$. It is of interest to note that logarithmic modes emerge in this solution.

We now examine the condition for which the metric is not conformally-Einstein.  Substituting the solution into (\ref{conformalcon}), we find that all of the components of $V_{\mu}$ vanish except for $V_r$.  We can solve for $V_r$, and substituting it into the remaining equations yields the constraint
\begin{equation}
c_0 + c_1 r =0\,.
\end{equation}
Thus, as long as $c_0$ and $c_1$ are not simultaneously zero, the metric is not conformally-Einstein.  On the other hand, if $c_0=c_1=0$, then we find that the metric $d\tilde s^2 = r^3 ds^2$ indeed becomes Ricci-flat.  Note that, as in the case of the homogeneous solution, the metric $d\hat s^2
= r ds^2$ describes a pp-wave in conformal gravity with the covariantly-constant null vector $\ell=\partial_v$.

\bigskip
\noindent{\bf Generalising solution 3:}
\be
ds^2 = \fft{dr^2}{r^2}+\fft{2du dv}{r}+dx^2+ r f du dx+ r^2 h du^2\,,
\ee
where
\bea
f &=& c_1+\fft{c_2 \log r}{r}+\fft{c_3}{r}+\fft{c_4}{r^2}\,,\nn\\
h &=& \ft{7}{36} c_1^2   + \fft{c_1(2c_3 -c_2)}{4r} + \fft{c_6}{r^2} +
\fft{c_8}{r^3}\cr
&& + \Big(\fft{c_1 c_2}{2r} + \fft{c_5}{r^2} +
\fft{c_7}{r^3}\Big)\log r + \fft{3 c_2^2 (\log r)^2}{8 r^2}\,.
\eea
The $c_i$'s are functions of $u$.  The metric is not conformal to an Einstein (Ricci-flat) metric provided that $c_1 c_2\ne 0$.  As in the case of the homogeneous solution, the metric $d\hat s^2 = r ds^2$ describes a pp-wave in conformal gravity.

\bigskip
\noindent{\bf Generalising solution 4:}
\be
ds^2 = \fft{dr^2}{r^2}+\fft{2du dv}{r}+\fft{dx^2}{r^4}+
    r^2 f du dx+r^8 h du^2\,,
\ee
where
\bea
f &=& c_1+\fft{c_2 \log r}{r^3}+\fft{c_3}{r^3}+\fft{c_4}{r^6}\,,\nn\\
h &=& \ft1{18} c_1^2  +\fft{c_1c_2}{12r^3}-\fft{c_2^2 (\log r)^2}{8r^6}+
 \fft{c_5}{r^6}\log r+\fft{c_6}{r^6}+\fft{c_7 \log r}{r^9}+
\fft{c_8}{r^9}+\fft{c_4^2}{4r^{12}}\,.
\eea
The $c_i$'s are functions of $u$.  Note that if $c_1=c_2=c_3=c_4=0$, then the solution reduces to the generalized solution 2. The metric is not conformally-Einstein provided that one of the $(c_1,c_2,c_5,c_6)$ is non-vanishing. As in the previous examples, the metric $d\hat s^2 = r ds^2$ describes a pp-wave in conformal gravity.

\bigskip
\noindent{\bf Generalising solution 5:}
\be
ds^2 = \fft{dr^2+ 2du dv+dx^2}{r^2}+f du dx+r^2 h du^2,
\ee
where
\bea
f &=& c_1 r+c_2+\fft{c_3}{r}+\fft{c_4}{r^2}\,,\nn\\
h &=& \ft18 c_1^2 r^2+\ft14 c_1 c_2 r+\ft18 (c_2^2+2c_1 c_3)+\fft{c_5}{ r}+\fft{c_6}{r^2}+\fft{c_7}{r^3}+\fft{c_8}{r^4}\,.
\eea
The $c_i$'s are functions of $u$. In this case, it is easier to describe 
the situations when the metric {\it is} conformally-Einstein: (i) $c_1=c_2=0$; (ii) $c_1\ne 0$ and $c_2=c_3=c_6=0$; (iii) $c_1 c_2\ne 0$ and 
$c_3=c_2^2/(3 c_1)$ and $c_5= c_1c_6/c_2+ c_2^3/(72 c_1)$. 
The metric
$d\hat s^2 = r^2 ds^2$ describes a pp-wave in conformal gravity.

Apparently none of these generalized solutions are characterized by their curvature invariants, which do not depend on the functions $f$ and $h$. Therefore, according to Theorem 2.3 in \cite{Coley:2009eb}, these metrics are all contained in the Kundt class. This implies that the global null vector admitted by each of these metrics is geodesic, expansion-free, shear-free and twist-free. In fact, these metrics are all written in the canonical Kundt form. Also, as previously mentioned, these metrics are all conformal to pp-waves, which are known to be of Kundt class.

\section{Bianchi IX metrics}

In this section, we consider a different class of metric ansatz: the cohomogeneity-one triaxial Bianchi IX metrics
\be
ds^2 = dr^2 + a^2\, \sigma_1^2 + b^2\, \sigma_2^2 + c^2\, \sigma_3^2\,,
\ee
where $a$, $b$ and $c$ are functions of $r$, and the $SU(2)$
left-invariant 1-forms $\sigma_i$ satisfy $d\sigma_i=-\ft12 \epsilon_{ijk}\,
\sigma_j\wedge\sigma_k$.   Substituting this into the equations of motion
of conformal gravity, we obtain a (rather complicated)
system of 4th-order differential equations for the functions
$a$, $b$ and $c$.  It can be verified that, in general, the necessary condition
(\ref{conformalcon}) for the metric to be conformally-Einstein cannot be
satisfied for any vector $V_\mu$ if one only imposes the aforementioned
equations of motion.  Thus, the vanishing of the Bach tensor does not in
general imply that the Bianchi IX metrics are conformally Einstein.

If we specialise to the case of {\it biaxial} Bianchi IX metrics by
setting, for example, $a(r)=b(r)$, then it turns out that the equations
following from the vanishing of the Bach tensor {\it do} now imply that
a solution to (\ref{conformalcon}) for a vector $V_\mu$ exists.
Furthermore, it can then
be explicitly verified that the equations following from requiring that
$\Omega^2\, ds^2$ be an Einstein metric imply that $\del_\mu\log\Omega$
is equal to the solution for $V_\mu$. In other words, for biaxial Bianchi IX metrics, the vanishing of the Bach tensor is sufficient to imply that the metric is conformally-Einstein, but this is no longer true for the case of triaxial Bianchi IX metrics.

Let us now present the biaxial case in some detail. Owing to the conformal symmetry, the most general ansatz can be expressed as
\begin{equation}
ds^2 = \fft{dr^2}{r^2 h} + h \sigma_3^2 + \sigma_1^2 + \sigma_2^2\,.
\end{equation}
The equations of motion reduce to
\begin{equation}
2r^4 h' h''' - r^4 h''^2 + 4 r^3 h' h'' - 4 r^2 h'^2 + 4 h^2 - 8 h +4=0\,,
\end{equation}
which can be solved analytically to give
\begin{equation}
h=1 + \fft{c_{-2}}{r^2} + \fft{c_{-1}}{r} + c_1 r  + c_2 r^2
\end{equation}
with
\begin{equation}
4c_{-2} \, c_2 = c_{-1} c_1\,.
\end{equation}
It is then straightforward to verify that the metric is conformally-Einstein.

For rotating black holes in the extremal limit, the near-horizon geometry is described by
a cohomogeneity-one metric whose components are functions of
the latitude coordinate $\theta$, with the level surfaces being $U(1)$
bundles over the AdS$_2$.  With an appropriate conformal transformation,
the metric takes the form
\begin{equation}
ds^2 = \fft{d\theta^2}{h} + h (d\phi + r dt)^2 + \fft{dr^2}{r^2} - r^2 dt^2\,,
\end{equation}
where
\begin{equation}
h= 1 + p_1 \cos\theta + p_2 \cos2\theta + q_1 \sin\theta + q_2 \sin2\theta\,,
\end{equation}
and
\begin{equation}
p_1^2 + q_1^2 = 4 (p_2^2 + q_2^2)\,.
\end{equation}
Note that the metric is simply a Lorentzian continuation of the
biaxial metric considered earlier. It is conformally Einstein, with
the vector $V_\mu$ in (\ref{conformalcon}) having the non-vanishing component
\begin{equation}
2V_\theta = -\fft{p_1 \cos\theta -2p_2 \cos 2\theta + q_1 \sin\theta
   -2q_2\sin 2\theta}{q_1\cos\theta + 2 q_2\cos 2\theta - p_1 \sin\theta
 - 2 p_2\sin 2\theta}
\end{equation}
This demonstrates that the near-horizon geometry of the most general
extremal black hole in conformal gravity is Einstein. This is a strong
indication that the rotating black hole obtained in \cite{Liu:2012xn} is
the most general black hole solution in conformal gravity, and it is
conformally Einstein.

\section{Homogeneous vacua in Einstein-Weyl gravity}

In this section, we consider solutions of Einstein-Weyl gravity with a cosmological constant:
\begin{equation}
e^{-1} {\cal L} = R - 2\Lambda + \ft12 \alpha C^{\mu\nu\rho\sigma}C_{\mu\nu\rho\sigma}\,.
\end{equation}
The linearized theory was studied in \cite{Lu:2011zk}.  It was shown that
the trace scalar mode decouples and hence the theory contains one massless
graviton and one ghost-like massive spin-2 mode.  There exists a special
point in the parameter space, known as critical gravity, for which the
massive spin-2 mode becomes
a logarithmic mode \cite{Lu:2011zk}.  New black holes, Lifshitz and
Schr\"odinger solutions can arise in Einstein-Weyl gravity \cite{Lu:2012xu}.
In this section, we show that more general classes of homogeneous
vacua (\ref{metric}) can also occur in Einstein-Weyl gravity.

\subsection{Diagonal metrics}

There are Lorentzian solutions given by
\be
ds^2=\ell^2\left( -r^{2z_3}dt^2+\fft{dr^2}{r^2}+r^{2z_1} dx_1^2+r^{2z_2} dx_2^2\right) \,,
\ee
for
\be
\Lambda = -\fft{1}{2\ell^2}\left( \sum_i z_i^2 +\sum_{i<j} z_iz_j\right)\,,\qquad
\alpha^{-1} = \fft{2}{3\ell^2} \left( \sum_i z_i^2-2\sum_{i<j} z_iz_j\right) \,,
\ee
where $i$ and $j$ are summed from 1 to 3.  Thus, we see that general
anisotropic Lifshitz solutions can arise in higher-derivative gravity
without the need for any matter energy-momentum tensor.

\subsection{Metrics with one off-diagonal term}

The first class of solutions with a single off-diagonal term in the metric are all Lorentzian signature and take the form
\be
ds^2=\ell^2\left( \fft{dr^2}{r^2}+r^{2z_1} dx_1^2+r^{2z_3} dx_3^2+r^{z_1+z_2} dx_1 dx_2\right) \,,
\ee
for
\bea
(1):&&z_2 = 2z_3-z_1\,,\qquad \Lambda=-\fft{3z_3^2}{\ell^2}\,,\qquad \alpha=\fft{\ell^2}{2z_1(z_3-2z_1)}\,,\nn\\
(2):&&z_1 = -\fft{z_3}{2}\,,\qquad \Lambda=-\fft{12z_2^2+4z_2z_3+11z_3^2}{32\ell^2}\,,\qquad \alpha=\fft{3\ell^2}{4z_3(z_3-z_2)}\,,\nn\\
(3):&& z_1= \fft{z_3}{4}\,,\qquad z_2=\fft{7z_3}{4}\,,\qquad \Lambda=-\fft{3z_3^2}{\ell^2}\,,\qquad \alpha=\fft{4\ell^2}{z_3^2}\,,\nn\\
(4):&& z_1 = -\fft{z_3}{2}\,,\qquad z_2=\fft{5z_3}{2}\,,\qquad \Lambda=-\fft{3z_3^2}{\ell^2}\,,\nn\\
(5):&& z_2 = -z_1\,,\qquad z_3=0\,,\qquad \Lambda=0\,,\qquad \alpha=-\fft{\ell^2}{4z_1^2}\,.
\eea
The second class of solutions can be of either Lorentzian or Euclidean
signature, depending on the parameters.  The metric for this class is given by
\be
ds^2=\ell^2\left( \fft{dr^2}{r^2}+r^{2z_1} dx_1^2+r^{2z_2} dx_2^2+r^{2z_3} dx_3^2+\sqrt{c_3} r^{z_1+z_2} dx_1 dx_2\right) \,,
\ee
with the constraints
\be
z_1 = -\fft{z_3}{2}\,,\qquad z_2=\fft{5z_3}{2}\,,\qquad \Lambda=\fft{3(11-2c_3)z_3^2}{2(c_3-4)\ell^2}\,,\qquad \alpha=\fft{(4-c_3)\ell^2}{4(4+5c_3)z_3^2}\,.
\ee
For $c_3>4$, the metric has Lorentzian signature, while for $c_3<4$ the metric has Euclidean signature.

In addition, we find  a Lorentzian solution
\be
ds^2=\ell^2\left( \fft{dr^2}{r^2}-r^{2z_1} dx_1^2+r^{2z_2} dx_2^2+r^{2z_3} dx_3^2+\sqrt{c_3} r^{z_1+z_2} dx_1 dx_2\right) \,,
\ee
for
\bea
z_1 &=& \fft14 \left( -2z_3+z_3\sqrt{3(c_3-4)}\right)\,,\qquad z_2 = \fft14 \left( -2z_3-z_3\sqrt{3(c_3-4)}\right)\,,\nn\\
\Lambda &=& -\fft{3c_3 z_3^2}{4(4+c_3)\ell^2}\,,\qquad \alpha=\fft{4(c_3+4)\ell^2}{(12-c_3)(3c_3-4)z_3^2}\,,
\eea
where the reality of the $z_i$ requires that $c_3>4$.

\subsection{Metrics with two off-diagonal terms}

We find two classes of solutions with two off-diagonal terms.  The first takes the form
\be
ds^2 = \ell^2\left( \fft{dr^2}{r^2}+r^{2z_2} dx_2^2+r^{z_1+z_2} dx_1 dx_2+r^{z_1+z_3} dx_1 dx_3\right)\,,
\ee
with
\bea
(1): && z_1 = -2z_2\,,\qquad z_3=4z_2\,,\qquad \Lambda=-\fft{3z_2^2}{\ell^2}\,,\qquad \alpha=-\fft{\ell^2}{20z_2^2}\,,\\
(2):&& z_1 = -\fft{z_2}{2}\,,\qquad z_3=\fft{5z_2}{2}\,,\qquad \Lambda=-\fft{3z_2^2}{\ell^2}\,,\qquad \alpha=\fft{4\ell^2}{z_2^2}\,,
\eea
and
\bea
(3):&&z_1 = \ft17 (\pm 3\sqrt{2}-2) z_2\,,\qquad z_3=\ft27 (2\mp 3\sqrt{2}) z_2\,,\nn\\
&&\Lambda = \fft{3(\pm 20\sqrt{2}-53)z_2^2}{196\ell^2}\,,\qquad \alpha=\fft{(\pm 2\sqrt{2}-1)\ell^2}{2z_2^2}\,.
\eea
In all three of these examples, the metric is Lorentzian.

The second class of solutions take the form
\be
ds^2 = \ell^2\left( \fft{dr^2}{r^2}+r^{2z_1} dx_1^2+r^{2z_2} dx_2^2+\sqrt{c_3} r^{z_1+z_2} dx_1 dx_2+r^{z_1+z_3} dx_1 dx_3\right)\,,
\ee
with
\bea
(4): &&z_1 = -2z_2\,,\qquad z_3=\fft{(4\pm 3\sqrt{2c_3}+c_3) z_2}{c_3-2}\,,\nn\\
&&\alpha = \fft{(c_3-2)\ell^2}{2(c_3\mp\sqrt{8c_3}-6)z_2^2}\,,\qquad
\Lambda=-\fft{3(48\pm 40\sqrt{2c_3}+10c_3\mp \sqrt{8c_3^3}+c_3^2)z_2^2}{8(c_3-2)^2\ell^2}\,.
\eea
and
\bea
(5): && z_1 = \fft{\sqrt{c_3}\left( 2\sqrt{c_3}\pm 3\sqrt{2(c_3-4)}\right) z_2}{36-7c_3}\,,\qquad z_3=-\fft{2\sqrt{c_3}\left(
2\sqrt{c_3}\pm3\sqrt{2(c_3-4)}\right) z_2}{36-7c_3}\,,\nn\\
&&\alpha = \fft{(7c_3-36)\ell^2}{2z_2^2 \left( (c_3-12)\mp \sqrt{8(c_3-4)c_3}\right)}\,,\nn\\
&&\Lambda = -\fft{3z_2^2 \left( (864-420c_3+53c_3^2)\pm 4(5c_3-18)\sqrt{2(c_3-4)c_3}\right)}{4(36-7c_3)^2 \ell^2}\,.
\eea
The metrics are Lorentzian also, provided that
$c_3$ is chosen appropriately so that all the parameters are real.

\section{Conclusions}

In this paper, we have constructed a variety of new solutions in conformal
gravity in four
dimensions.  Quadratic gravities in four dimensions are special in that
Einstein metrics are automatically solutions.  In conformal gravity,
Einstein metrics with an arbitrary conformal factor are also solutions.
Indeed, the most general spherically-symmetric black holes in conformal
gravity can be shown to be locally conformal to the Schwarzschild-AdS
black hole.  The most general metric in conformal gravity in the Plebanski
class is also locally conformal to the (neutral) Plebanski-Demianski metric,
which is Einstein.  It is of interest to look for
 solutions that are not merely conformal scalings of Einstein metrics.
Such solutions are known to exist, although there are few
explicit examples.

   We have constructed five new examples of homogeneous vacua solutions
that are not conformally Einstein.  These metrics exhibit generalized
Lifshitz anisotropic scaling.  We also studied
their Petrov classification.  It turns out that four of the
solutions are conformal to pp-waves (defined to be metrics
admitting a covariantly-constant null vector).  We were able to
generalize these solutions further, to obtain inhomogeneous metrics with
multiple independent free functions.

It is of interest to compare conformal gravities in four and six dimensions.  In six dimensions, there can be three independent conformal invariant terms and only one specific combination admits Einstein metrics as solutions \cite{Bastianelli:2000rs,Metsaev:2010kp,Oliva:2010zd,Lu:2011ks}. Thus the general solutions in conformal gravities in six dimensions are not conformally Einstein.  Furthermore, even for the one that admits Einstein metrics, the study of the spherically-symmetric solutions shows that there exist classes of solutions that are not conformally Einstein \cite{Lu:2013hx}.

  We have also obtained new homogeneous vacua with generalized multiple
anisotropic Lifshitz scaling symmetry in Einstein-Weyl gravity.  Our
results show that there exist wide classes of homogeneous vacua in
higher-derivative gravities.  Their classification and application to
non-relativistic holography require further investigation.

\section*{Acknowledgements}

We are grateful to David Chow, Maciej Dunajski, James Liu and Yi Pang for helpful
conversations. J.F.V.P. is grateful to Beijing Normal University for hospitality during the initial stages of this work. The research of L\"u is supported in part by the NSFC grants 11175269 and 11235003. The research of C.N.P. is supported in part by DOE grant DE-FG03-95ER40917. The research of J.F.V.P. is supported in part by NSF grant PHY-0969482.

\appendix

\section{Petrov Classification}

  Here, we summarise some key aspects of the Petrov classification of the
Weyl tensor.  This amounts to a classification of the structure of the
eigenvalues $\lambda$
of the equation $\ft12 C_{\mu\nu\rho\sigma}\, X^{\rho\sigma}=
\lambda\, X_{\mu\nu}$.  It can be restated as follows.
 We begin by defining the tensor
\be
Q_{\mu\nu\rho\sigma}= \ft12 (C_{\mu\nu\rho\sigma} -
   \im\, ^*C_{\mu\nu\rho\sigma} -\im\, C^*_{\mu\nu\rho\sigma} -
    {^*C}^*_{\mu\nu\rho\sigma})\,,
\ee
where the left and right duals of the Weyl tensor are defined by
\be
^*C_{\mu\nu\rho\sigma}= \ft12 \ep_{\mu\nu\alpha\beta}\,
     C^{\alpha\beta}{}_{\rho\sigma}\,,\qquad
C^*_{\mu\nu\rho\sigma}= \ft12 \ep_{\rho\sigma\alpha\beta}\,
  C_{\mu\nu}{}^{\alpha\beta}\,.
\ee
Quadratic and cubic curvature invariants $I$ and $J$ are then defined by
\be
I= \ft1{16}\,Q_{\mu\alpha\nu\beta}\, Q^{\nu\alpha\mu\beta}\,,\qquad
J= \fft1{192}\, Q^{\mu\nu\rho\sigma}\, Q_{\mu\alpha\rho\beta}\,
  Q_{\nu}{}^\alpha{}_\sigma{}^\beta\,.
\ee

   If the spacetime is conformally flat, $C_{\mu\nu\rho\sigma}=0$, then it
is of Petrov type O.  For all other spacetimes, we have the possibilities:
\bea
I^3\ne 27 J^2:&& \qquad\qquad \hbox{Type I}\,,\nn\\
&&\nn\\
I^3=27 J^2\ne 0:&&\qquad \qquad \hbox{Type II or Type D}\,,\nn\\
&&\nn\\
I=J=0:&&\qquad\qquad \hbox{Type N or Type III}
\eea

  The six distinct Petrov types arise as follows.  At the top, with
$I^3\ne 27 J^2$, is Type I. In the second level, where $I^3=27J^2\ne 0$,
lie Type II and Type D.  At the lowest level, where $I=J=0$, lie
Type III, Type N and Type O:
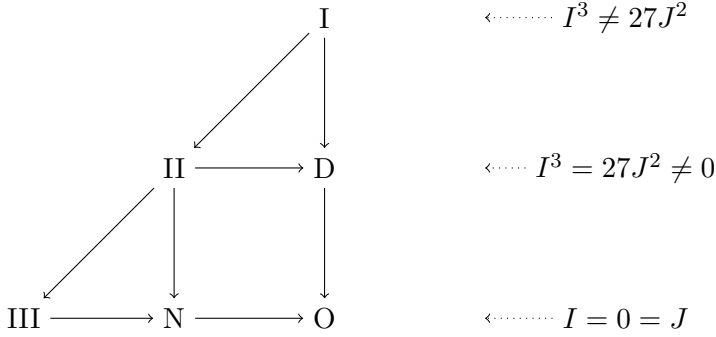
\bigskip

\begin{tikzpicture}[node distance=2cm, auto]
\node (I) {I};
  \node (D) [below of=I] {D} ;
  \node (II) [left of=D] {II};
  \node (O) [below of=D] {O};
  \node (N) [below of=II] {N};
  \node (III) [left of=N] {III};
  \node (Is) [right of=I] {};
  \node (Ds) [right of=D] {};
  \node (Os) [right of=O] {};
  \node (L1) [right of=Is] {$I^3\ne 27 J^2$};
  \node (L2) [right of=Ds] {$I^3=27 J^2 \ne 0$};
  \node (L3) [right of=Os] {$I=0=J$};
  \draw[->] (I) to node {}  (II);
  \draw[->] (I) to node  {} (D);
  \draw[->] (II) to node [swap] {} (D);
  \draw[->] (D) to node {} (O);
  \draw[->] (II) to node {} (N);
  \draw[->] (II) to node {} (III);
  \draw[->] (III) to node {} (N);
  \draw[->] (N) to node {} (O);
  \draw[->,dotted] (L1) to node {} (Is);
  \draw[->,dotted] (L2) to node {} (Ds);
  \draw[->,dotted] (L3) to node {} (Os);
\end{tikzpicture}
\bigskip

Type II can be distinguished from Type D by the fact that, in a Type II
spacetime, there exists a non-vanishing vector $K^\mu$ such that
\be
C_{\mu\nu\rho\sigma}\, K^\mu\, K^\rho = a\, K_\nu\, K_\sigma\,,\qquad
^*C_{\mu\nu\rho\sigma}\, K^\mu\, K^\rho = b\, K_\nu\, K_\sigma\,,
\ee
for some $a$ and $b$.

Type O is conformally flat, with $C_{\mu\nu\rho\sigma}=0$.
Type N can be distinguished from Type III by the fact that in a Type N
spacetime there exists a non-vanishing vector such that
\be
C_{\mu\nu\rho\sigma}\, K^\sigma=0\,.
\ee

The orders of specialisation from one Petrov type to another are
indicated by the arrows in the diagram.

\end{document}